\documentclass[12pt]{article}
\usepackage{graphics}

\def\prb#1#2#3{Phys. Rev. B \textbf{#1}, #2 (#3)}

\def\be{\begin{equation}}
\def\ee{\end{equation}}
\def\bfi{\begin{figure}}
\def\efi{\end{figure}}
\def\bea{\begin{eqnarray}}
\def\eea{\end{eqnarray}}

\begin{document}

\begin{center}
{\Large \bf Growth Law and Superuniversality in the Coarsening of Disordered Ferromagnets}
\vskip1cm
F.Corberi$^1$, E.Lippiello$^2$, A.Mukherjee$^3$, S.Puri$^3$ and M.Zannetti$^1$ \\
\vskip0.5cm
$^1$Dipartimento di Matematica e Informatica and CNISM, Unit\`a di Salerno,
Universit\`a di Salerno, via Ponte don Melillo, 84084 Fisciano (SA), Italy. \\
$^2$Dipartimento di Scienze Ambientali, Seconda Universit\`a di Napoli, \\
Via Vivaldi, Caserta, Italy. \\
$^3$School of Physical Sciences, Jawaharlal Nehru University, \\
New Delhi--110067, India. \\
\end{center}

\begin{abstract}
We present comprehensive numerical results for domain growth in the two-dimensional {\it Random Bond Ising Model} 
(RBIM) with nonconserved Glauber kinetics. We characterize the evolution via the {\it domain growth law}, 
and two-time quantities like the {\it autocorrelation function} and {\it autoresponse function}. 
Our results show evidence for the crossover from a pre-asymptotic 
regime with ``power-law growth with a disorder-dependent exponent'' to an asymptotic regime 
with ``logarithmic growth''. We compare this behavior with previous results on one-dimensional
disordered systems and we propose a unifying picture in a renormalization group framework. 
We also study the corresponding crossover in the scaling functions for the two-time quantities.
Super-universality is found not to hold. Clear evidence supporting the dimensionality dependence
of the scaling exponent of the autoresponse function is obtained.
\end{abstract}

\newpage

\section{Introduction}

The {\it coarsening dynamics} or {\it domain growth} of a thermodynamically unstable system, such as a 
ferromagnet or a binary mixture quenched below the critical point, is a well-established paradigm of slow 
relaxation. From the pioneering work of Marro et al.~\cite{MLK} up to the reviews of Bray~\cite{Bray} 
and Puri~\cite{Puri}, the equal-time properties have been extensively investigated. Recently, the interest 
has shifted to the study of two-time quantities, since the ordering process offers a simple and non-trivial 
arena for the understanding of aging phenomena~\cite{reviews,chpt}. There is also a long-standing interest 
in studying the robustness of the basic features of coarsening systems when quenched disorder 
(without frustration) is introduced \cite{pcp91,sp04}. Taking the Ising model with ferromagnetic couplings
as the basic system, 
the class of disordered systems investigated includes 
the {\it random field Ising model} (RFIM)~\cite{nattermann} and the {\it random bond Ising model} (RBIM)~\cite{sp04}.

The first effect of disorder is to change the relaxation mechanism from purely
energy lowering to activated, due to the creation of energy barriers~\cite{Lai}. The
question is to what extent this affects the growth law, namely
the dependence on time of the linear size of domains $L(t)$.
Intuitively, the presence of energy barriers is expected to slow down the coarsening process. 
However, the problem turns out not to be so straightforward, since the barrier size may 
depend on the domain size, thus determining the nature of the asymptotic growth law through a non-trivial 
feedback process. So far, this issue has been well understood in the case of one-dimensional systems. 
The $d=1$ RFIM~\cite{decandia} and the $d=1$ RBIM~\cite{EPL} offer instructive and clear instances of 
disorder forming either $L$-dependent barriers (in the RFIM) or $L$-independent barriers (in the RBIM). 
In both cases, there is a crossover. In the $d=1$ RFIM, it occurs from pure-like power-law growth to 
disorder-dominated logarithmic growth. In the $d=1$ RBIM, the crossover takes place in the reverse order: 
from a pre-asymptotic disorder-dependent power law to an asymptotic pure-like power-law growth.

In higher dimensions, a crossover from algebraic to logarithmic behavior has been found in the growth 
law of the RFIM~\cite{pp93,eo90,rao,aron}. A crossover from algebraic to slower than power-law
behavior, possibly logarithmic, has been reported also in the case of the random site 
Ising model (RSIM)~\cite{Grest,Lai,Park}. According to the phenomenological theory of
Huse and Henley (HH)~\cite{HH}, this means that in these systems energy barriers ought to scale like a power
of $L$. 

Instead, in the $d=2$ RBIM the role of barriers and the nature of the growth law has 
remained controversial, since the available numerical evidence shows a steady algebraic 
growth with a disorder-dependent dynamical exponent~\cite{ppr04,HP1,HP2}. 
The open issue is whether the data manage to probe the genuine late time behavior,
in which case energy barriers ought to scale logarithmically with $L$~\cite{ppr04},
or the HH paradigma is valid also for this system, in which case the data would be
probing a long pre-asymptotic regime, 
eventually to be followed by a crossover to logarithmic growth~\cite{Cuglia,Iguain}. 
That an HH-type scenario might yield an effective algebraic law at intermediate times
had already been noticed by Bouchaud et al.~\cite{Bouchaud} in the context of spin glasses.
Our contribution in this paper is precisely the clearing up of this point, producing the 
numerical evidence for the existence of the crossover to logarithmic growth in the $d=2$ RBIM.

A related important question is how disorder affects the scaling functions,
namely whether it enters only through the modified growth law or if there is also an 
explicit disorder dependence. In the first case, scaling functions
ought to be the same for pure and disordered 
systems, once time is reparametrized through $L(t)$. This is referred to as 
the {\it super-universality} (SU) hypothesis~\cite{su}, whose validity has been supported 
mainly from numerical results for the equal-time correlation function 
or structure factor~\cite{pcp91,rao,aron,ppr04}.
More recently, the validity of SU has been extended by Sicilia et al.~\cite{Sicilia} 
to the geometrical properties of domain structures and by
Henkel and Pleimling (HP)~\cite{HP1,HP2} 
to two-time quantities, such as
the order parameter autocorrelation and autoresponse functions in the $d=2$ RBIM.
However, next to the features supporting SU validity,
there exists also some evidence showing SU violation, as reported by
HP~\cite{HP2} in the short distance behavior of the $d=2$ RBIM
and by Park and Pleimling~\cite{Park} in the space-time correlation function in the
RSIM. What is missing, here, is the general framework where to fit 
coherently these conflicting pieces of information, with the end-result that
the issue of SU validity is regarded as yet to be understood.

By contrast, a clear picture emerges from 
the $d=1$ cases mentioned above~\cite{decandia,EPL}, 
where the crossover can be investigated with considerable precision and turns out not 
to be limited to the growth law, but to take place also in the correlation and
response functions. Then, the issue of SU validity or lack of validity is
reduced to deciding whether the asymptotics are or are not disorder-dominated.
Adopting this point of view, our finding of the crossover in the growth law in $d=2$ RBIM
strongly suggests that in this system too there might be full violation of SU 
and that this might fit into the pattern extracted from the $d=1$ systems.
Indeed, this is what we find from the autocorrelation  
and autoresponse function data 
and, putting together these new data with those for $d=1$ systems, 
we have succeded in constructing a new and coherent picture of general character for the coarsening dynamics of 
weakly disordered systems. This is most effectively formulated
into a renormalization group framework 
in terms of the competition between a {\it pure fixed point} and a {\it disordered fixed point}. 
Then, the variety of different scenarios observed arises from the different possible choices in the 
relative stability of these fixed points.

Finally, the presence of disorder allows for an exhaustive investigation of the scaling properties of the 
dynamical susceptibility. The data in the pre-asymptotic regime of the $d=2$ RBIM, where the growth 
law is algebraic with good precision, together with those for the $d=1$ RBIM, produce conclusive 
evidence for the $d$-dependence of the exponent involved. Our results are in agreement with the
conjecture put forward in studies of the pure system~\cite{chpt,single,CLZ,generic}. 
A preliminary account of this result has been presented in Ref.~\cite{EPL}.

This paper is organized as follows. In Sec.~\ref{GF}, we provide a general framework, based
on renormalization group ideas, to 
understand the effects of disorder on domain growth. In Sec.~\ref{d1}, we summarize our 
understanding of coarsening in $d=1$ disordered systems. The bulk of the paper is contained 
in Sec.~\ref{d2}, where we present comprehensive numerical results for domain growth in the $d=2$ RBIM. 
We conclude the paper with a summary and discussion in Sec.~\ref{conc}.

\section{General Framework}
\label{GF}

This section is devoted to an overview, using scaling arguments and renormalization group (RG)
terminology, of the possible deviations from pure-like behavior due to disorder. 

\subsection{Growth Law}
\label{GL}

The linear size of domains in pure systems asymptotically grows according to the power law
\be
L(t) = D_0 \, t^{1/z} ,
\label{GR.1}
\ee
which holds when time $t$ is large enough for $L(t)$ to dominate all other lengths in the problem. 
Here, $D_0$ is a prefactor which is dependent on temperature, and $z$ is the dynamical 
exponent.
It is straightforward to verify that the above power law satisfies the homogeneity relation
\be
L(t) = l L\left ( {t \over l^z} \right )
\label{ho.1}
\ee
where $l$ is arbitrary and can be regarded as the scale parameter in an RG
transformation. In this paper we shall consider only non-conserved dynamics. In pure systems
the dynamical exponent is $z=2$, independent of the system dimensionality~\cite{Bray}.
 
Assuming scaling, the growth law for disordered systems is obtained generalizing
Eq.~(\ref{ho.1}) to
\be
L(t,\epsilon) = l {\cal L}\left ( {t \over l^z} , {\epsilon \over l^{\phi}} \right )
\label{ho.2}
\ee
where $\epsilon$ is a parameter quantifying the amount of disorder and $\phi$
is the corresponding scaling exponent. Once the growth law is written in this form,
the standard crossover scenario follows from the competition between the pure fixed
point at $\epsilon =0$ and the disordered fixed point at $\epsilon =\infty$.
The relative stability of these two fixed points is regulated by the sign of
$\phi$, with the pure fixed point being stable or unstable for $\phi > 0$ 
or $\phi < 0$, respectively.
In the first case the asymptotic
growth law would remain the pure one, while, in the second case, there would
be the crossover to an asymptotic behavior different from that of Eq.~(\ref{ho.1}) and
controlled by the disordered fixed point.

In place of investigating directly the sign of $\phi$, 
it turns out to be more convenient to introduce the crossover length $\lambda(\epsilon) \sim \epsilon^{1/\phi}$ 
and to rewrite Eq.~(\ref{ho.2}) as\footnote{As we shall see, in the case of the
$d=1$ RBIM Eq.~(\ref{ho.3}) applies also to the more general case where $\lambda(\epsilon)$
does not scale like a power of $\epsilon$.}
\be
L(t,\epsilon) = l {\cal L}^{\prime} \left ( {t \over l^z} , {\lambda(\epsilon) \over l} \right ).
\label{ho.3}
\ee
At fixed points the crossover length values are either $\lambda^*=0$ or $\lambda^*=\infty$, and
since $\lambda(\epsilon)$ flows from $\infty$
toward $0$, the fixed point corresponding to $\lambda^*=0$ is stable, while the one with
$\lambda^*=\infty$ is unstable. 
Therefore, the relevancy or irrelevancy of disorder can be deduced from the behavior
of $\lambda(\epsilon)$ as $\epsilon \to 0$. Specifically, since $\epsilon = 0$ corresponds to
the pure fixed point and the two possible limits  of $\lambda(\epsilon)$ are
\be
\lim_{\epsilon \to 0} \lambda(\epsilon)    = \left \{ \begin{array}{ll}
        0  \\
        \infty
        \end{array}
        \right .
        \label{I.4}
        \ee
if the first one applies, the pure fixed point is stable, while 
if the second limit applies, the disordered fixed point is 
stable.

As a final remark, notice that making the choice $l=t^{1/z}$ in Eq.~(\ref{ho.3}), 
the growth law takes the standard scaling form
\be
L(t,\epsilon) = t^{1/z} \widetilde{{\cal L}}\left ( {\lambda(\epsilon) \over t^{1/z}} \right ).
\label{GR.4}
\ee  

\subsection{Two-time Quantities}

Once the role of $\lambda(\epsilon)$ is understood in the growth law, the next step is to 
establish how it affects the other observables in the problem. 

In this paper, we shall be concerned with spin systems which are quenched below the critical 
temperature at $t=0$. For what follows, the basic two-time quantities are the {\it autocorrelation function} 
and the {\it autoresponse function}. The first of these is defined by
\be
C(t,t_w) = \langle \sigma_i(t) \sigma_i(t_w) \rangle -  \langle \sigma_i(t) \rangle \langle \sigma_i(t_w)\rangle .
\label{AUCF}
\ee
Here, $\sigma_i =\pm 1$ is the spin variable at the lattice site $i$, $(t,t_w)$ are a pair
of times after the quench, with the convention $t_w \leq t$, and angular brackets $\langle \cdot \rangle$
denote average over all sources of randomness in the problem: initial condition, thermal fluctuations
and quenched disorder. After taking these averages $C(t,t_w)$ is 
space-translationally invariant and there is no dependence on the choice of $i$.

The autoresponse function is defined by
\be
R(t,t_w) = \left. {\delta \langle \sigma_i(t) \rangle_h \over \delta h_i(t_w)} \right |_{h=0} ,
\label{IAR.1.01}
\ee
where $h_i(t)$ is a space and time-dependent external field conjugate to $\sigma_i$, and
$\langle \cdot \rangle_h$ denotes the average in the presence of the field. This {\it instantaneous} 
response function is very difficult to measure, both numerically and experimentally. For this reason, 
time integration is used as a means to improve the signal to noise ratio~\cite{respfunct}. We shall 
focus on the {\it zero-field-cooled susceptibility} (ZFCS), obtained by switching on the external field 
from the time $t_w \geq 0$ up to the observation time $t$:
\be
\chi(t,t_w) = \int_{t_w}^t \, dt^{\prime} \, R(t,t^{\prime}).
\label{ZFC}
\ee
The motivations and the advantages for using this form of integrated response function, with respect
to others, have been discussed at length in Ref.~\cite{CLZ}.

Let us denote by ${\cal O}(t,t_w)$ either one of the above observables. In the pure case 
and for quench temperatures low enough to neglect thermal fluctuations, the {\it aging-scaling} relation 
is obeyed~\cite{reviews,chpt,furukawa}
\be
{\cal O}(t,t_w) = L^{-\alpha}(t_w)F \left ({L(t) \over L(t_w)} \right ) .
\label{IAF.1}
\ee
where $\alpha$ is the scaling exponent and $F(x)$ is the scaling function. 

The extension of Eq.~(\ref{IAF.1}) to the disordered case is a non-trivial problem. 
According to the SU hypothesis, disorder should affect only the growth law, leaving 
the form of Eq.~(\ref{IAF.1}) unaltered with the same scaling function, that is
\be
{\cal O}(t,t_w,\epsilon) = L^{-\alpha}(t_w,\epsilon)F \left ({L(t,\epsilon) \over L(t_w,\epsilon)} \right ) .
\label{IAF.1bis}
\ee 
In other words, once time 
is re-parametrized by $L(t,\epsilon)$, there should be no difference between disordered and pure systems. 
This is a strong statement.

Alternatively, one can allow for an explicit disorder-dependence of the scaling function
\be
{\cal O}(t,t_w,\epsilon) =  L_w^{-\alpha}\widetilde{F} \left ({L \over L_w}, 
{\lambda \over L_w} \right ) ,
\label{IAF.4}
\ee
and extend the crossover scenario to two-time observables. Here
$L(t,\epsilon)$ has been denoted by $L$, $L(t_w,\epsilon)$ by $L_w$
and $\lambda(\epsilon)$ by $\lambda$.  
Using the notation $x=L/L_w$ and $y=\lambda/L_w$, 
the $x \ll y$ region is controlled by the unstable $y^*=\infty$ fixed point, 
while the $x \gg y$ region is controlled by the stable $y^*=0$ fixed point. 
Adopting the convention that hat-free symbols refer to the pure fixed point and 
hatted symbols to the disordered fixed point, if disorder is an irrelevant 
perturbation from the analysis presented in the previous subsection follows that the
stable fixed point $y^*=0$ is pure and that
the general scaling relation in Eq.~(\ref{IAF.4}) yields the limiting behaviors
\be
L_w^{-\alpha}\widetilde{F}(x,y)   \simeq \left \{ \begin{array}{ll}
        L^{-\widehat{\alpha}}_w \widehat{F}(x),\;\; $for$ \;\; x \ll y , \\
        L_w^{-\alpha} F(x),\;\; $for$ \;\; x \gg y ,
        \end{array}
        \right .
        \label{crossover2}
        \ee
where $\widehat{F}(x) \neq F(x)$. The exponent $\widehat{\alpha}$ is different 
from or equal to $\alpha$, depending on whether $\lim_{y \to \infty}\widetilde{F}(x,y)$ is 
singular or not. Conversely, if disorder is a relevant perturbation
the stable fixed point $y^*=0$ is disorder-controlled and the limiting behaviors are reversed
\be
L_w^{-\alpha}\widetilde{F}(x,y)   \simeq \left \{ \begin{array}{ll}
        L_w^{-\alpha} F(x),\;\; $for$ \;\; x \ll y , \\
        L^{-\widehat{\alpha}}_w \widehat{F}(x),\;\; $for$ \;\; x \gg y 
        \end{array}
        \right .
        \label{crossover}
        \ee
with $\widehat{\alpha}$ being different from or equal to $\alpha$, 
depending on whether $\lim_{y \to 0}\widetilde{F}(x,y)$ is singular or not.
For instance, in the case of the autocorrelation 
function there are no singularities in $\widetilde{F}(x,y)$, since $\alpha_C = \widehat{\alpha}_C = 0$ 
as a consequence of the compact nature of domains, both for pure and disordered systems.

In the general framework outlined above, the SU hypothesis holds as an asymptotic statement
when disorder is irrelevant, while SU is violated and the asymptotics of disordered and 
pure systems are different when disorder is relevant. 

\section{Domain Growth in $d=1$ Disordered Ising Systems}
\label{d1}

Let us now discuss $d=1$ disordered systems. These will provide clear realizations of the 
possible scenarios outlined in the previous section. As a general comment, it should be kept in mind
that even if $d=1$ systems are paramagnetic at any finite temperature, as long as the domain
size $L(t)$ is smaller than the equilibrium correlation length $\xi(T)$, the coarsening dynamics
is the same as in the quench to $T=0$ where ferromagnetic ordering occurs. This is explained
in detail in Refs.~\cite{decandia,EPL,anal1}.

We consider the ferromagnetic Ising chain with Hamiltonian
\be
{\cal H} = -\sum_{i=1}^{N-1} J_i \sigma_i\sigma_{i+1} -\sum_{i=1}^N h_i\sigma_i, \quad \sigma_i = \pm 1 .  
\label{GL.2}
\ee
The system evolves with non-conserved Glauber spin-flip dynamics \cite{Puri}. Interfaces in this case are 
point defects, separating domains of up spins from domains of down spins. In order to identify the 
length $\lambda$, and to keep the discussion simple, let us focus on the energetics of the particular 
configuration $[\sigma]_j$, containing a single defect breaking the bond $J_j$:
\be
\sigma_i = \left \{ \begin{array}{ll}
        1, \;\;$for$ \;\; i \leq j , \\
        -1,  \;\; $for$ \;\; i > j.
        \end{array}
        \right .
        \label{1d.1}
        \ee
The energy change when the defect moves from the bond $j$ to another location $r$ lattice spacings away with,
say, $r > 0$, is given by
\be
\Delta E_{j,j+r} = 2[J_{j+r} - J_{j} - \sum_{i=j}^{j+r}h_i].
\label{1d.2}
\ee
In the pure system, the ferromagnetic coupling is the same for all pairs of nearest neighbors ($J_i=J>0$). 
In the absence of an external field ($h_i = 0$), $\Delta E_{j,j+r} = 0$ for any $r$. The potential landscape is 
flat and defects perform free diffusion, yielding the behavior of Eq.~(\ref{GR.1}) in the large-time regime~\cite{anal1}. 

In the $d=1$ ferromagnetic RBIM, the couplings $J_{i}$ are non-negative independent random variables, and $h_i = 0$ 
as in the pure case. For a given configuration of the couplings, there are now local minima and maxima in the 
potential landscape. A local minimum occurs when a defect breaks a bond flanked, to the right and 
to the left, by stronger bonds. Similarly, a defect sits on a local maximum when the opposite bond 
configuration occurs. The energy difference in this case is given by
\be
\Delta E_{j,j+r} = 2(J_{j+r} - J_{j}) ,
\label{1d.3}
\ee
which depends on the actual values of the couplings involved, but not on their distance. 
For instance, taking the $J_{i}$ to be uniformly distributed in the interval $(1-\epsilon,1+\epsilon)$ 
with $0 \leq \epsilon \leq 1$, the upper bound on $\Delta E_{j,j+r}$ is given by $4\epsilon$. 
Using the Arrhenius-type relation between escape time and barrier height
$t_B \sim e^{E_B/T}$, there remains defined a characteristic time
\be
\tau \sim  e^{4\epsilon/T} ,
\label{EB.2}
\ee 
as the time needed to overcome the highest energy barrier, and a characteristic length scale
\be
\lambda \sim \left [e^{4\epsilon/(zT)} - 1 \right ].
\label{1d.4}
\ee
We take the Boltzmann constant $k_B=1$ throughout the paper.

In the $d=1$ RFIM, the ferromagnetic coupling is the same for all pairs of nearest neighbors ($J_i=J>0$), 
as in the pure case, while $h_i=\pm \epsilon$ is an uncorrelated random field with 
expectations\footnote{Clearly, here the average is only over the quenched randomness.} 
$\langle h_i \rangle =0$ and $\langle h_ih_j \rangle =\epsilon^2\delta_{ij}$. In this case, 
domain walls perform random walks in a random potential of the Sinai type~\cite{FLDM}. The 
landscape contains minima and maxima, as in the $d=1$ RBIM. However, the crucial difference is 
that the height of energy barriers is not bounded and scales with the distance traveled by the defect:
\be
\Delta E_{j,j+r} = - \sum_{i=j}^{j+r}h_i \sim \epsilon r^{1/2}.
\label{EB.4}
\ee
The characteristic disorder length is obtained by balancing thermal energy with the barrier height 
\be
\lambda = (T/\epsilon)^2.
\label{EB.4bis}
\ee

Therefore, from Eqs.~(\ref{1d.4}) and (\ref{EB.4bis}), it follows that 
when $\epsilon \rightarrow 0$ the two possible scenarios, originating 
from Eq.~(\ref{I.4}) and analyzed in detail in Sec.~\ref{GF}, find 
realization in the $d=1$ RBIM and RFIM, respectively. Indeed, this is what has been found~\cite{decandia,EPL}.
In the $d=1$ RBIM disorder is irrelevant since from Eq.~(\ref{1d.4}) follows $\lim_{\epsilon \to 0}\lambda =0$.  
Our numerical simulations~\cite{EPL} have shown very precisely that the growth law, the 
autocorrelation function and the ZFCS scale according to Eqs.~(\ref{GR.4}), (\ref{IAF.4}) 
and (\ref{crossover2}), with a crossover from pre-asymptotic disordered to asymptotic pure-like 
behavior. Actually, in the pre-asymptotic regime, the growth law is algebraic with a disorder-dependent 
growth exponent, indicating that $\lambda$ should more properly be regarded as a marginal 
scaling field. A similar behavior will be encountered in the $d=2$ RBIM
discussed in Sec.~\ref{d2}. Furthermore, the ZFCS exponent vanishes:
\be
\alpha_{\chi}=0
\label{alpha}
\ee
for all disorder values, as in the pure case~\cite{anal1}. Hence, the ZFCS scaling function, 
like that of the autocorrelation function, does not have singularities for $y \rightarrow \infty$.

On the other hand, the $d=1$ RFIM~\cite{decandia} is a realization of the opposite case, since 
Eq.~(\ref{EB.4bis}) shows that $\lim_{\epsilon \to 0}\lambda =\infty$ 
and, therefore, that disorder is  relevant. The growth law, the 
autocorrelation function and ZFCS scale according to Eqs.~(\ref{GR.4}), (\ref{IAF.4}) 
and (\ref{crossover}) and the crossover occurs from pure-like to disordered behavior. 
Again, as in the $d=1$ RBIM, the exponents of the autocorrelation function and ZFCS vanish in 
the pure and disordered systems, revealing the absence of singularities in the scaling functions as $y \rightarrow 0$.

Because of its importance in this paper, it is useful to recall the result for the growth law in 
the $d=1$ RFIM. The limiting forms of the scaling function in Eq.~(\ref{GR.4}) 
are given by \cite{decandia}
\be
{\cal L}(y)  \sim \left \{ \begin{array}{ll}
        y(\ln y)^2, \;\;$for$ \;\; y \ll 1 , \\
        D,  \;\; $for$ \;\; y \gg 1.
        \end{array}
        \right .
        \label{EB.6}
        \ee 
Thus, in the pre-asymptotic regime (where $y \gg 1$ or $L(t) \ll \lambda$), the thermal energy 
exceeds the barrier height and growth takes place via free diffusion, as in the pure system 
(there is a possible dependence of the diffusion constant $D$ on disorder.) 
In the asymptotic regime (where $y \ll 1$ or $L(t) \gg \lambda$), diffusion is of the 
Sinai type with $L(t) \sim  (\ln t)^2$~\cite{Sinai}. This is a consequence of the Arrhenius law  
with a barrier height which depends on $L$ according to  Eq.~(\ref{EB.4}). Then, 
the overall growth law, which interpolates between limiting behaviors and accounts for Eq.~(\ref{EB.6}), can be written as
\be
t \simeq L^z e^{(L/\lambda)^{1/2}}.
\label{Arr.1}
\ee

\section{Domain Growth in the $d=2$ Random Bond Ising Model}
\label{d2}

Let us next consider the $d=2$ RBIM in the absence of an external field, with Hamiltonian
\be
{\cal H} = -\sum_{\langle ij \rangle} J_{ij}\sigma_i\sigma_{j} ,
\ee
where the sum runs over nearest-neighbor pairs. The ferromagnetic couplings $J_{ij}$ are independent 
random variables with the same statistics as in the $d=1$ case, i.e., uniformly distributed in the interval
$(1-\epsilon,1+\epsilon)$ with $0 \leq \epsilon \leq 1$. 

Let us first summarize the preexisting theoretical and numerical situation.
According to the phenomenological HH~\cite{HH} theory, energy barriers in this system (when $d>1$) 
ought to scale as a power of the domain size:
\be
E_B(L) = \kappa L^{\psi} ,
\label{GL.2hh}
\ee
where $\kappa$ is a disorder-dependent prefactor such that 
\be
\lim_{\epsilon \to 0}\kappa = 0 
\label{GL.2tris}
\ee
and $\psi$ is an exponent dependent on dimensionality with the value $\psi = 1/4$ for $d=2$.
 
Let us, then, see what are the consequences on the basis of the general considerations made above.
Identifying the characteristic 
disorder scale through the matching of thermal energy with barrier height, one has
\be
\lambda = (T/\kappa)^{1/\psi} 
\label{GL.2bis}
\ee
which implies $\lim_{\epsilon \to 0}\lambda = \infty$. Therefore,
in the HH theory disorder is relevant, which implies
that the scenario found in the $d=1$ RFIM should be replicated  
with a diffusive pre-asymptotic regime 
followed by asymptotic logarithmic growth
\be
L(t)  \sim \left \{ \begin{array}{ll}
        t^{1/z}, \;\;$for$ \;\; L \ll \lambda , \\
        (\ln t)^{1/\psi},  \;\; $for$ \;\; L \gg \lambda
        \end{array}
        \right .
        \label{GL.4}
        \ee
and with an interpolating formula analogous to Eq.~(\ref{Arr.1}) 
\be
t \simeq L^z e^{(L/\lambda)^{\psi}}.
\label{GL.3}
\ee

However, this HH scenario has never been actually observed. 
Rather, from a number of experiments on random magnets~\cite{exp} and from extensive 
simulations of the $d=2$ RBIM~\cite{ppr04,HP1,HP2}, the observed growth seems to be well-described by an algebraic law
\be
L(t) \sim t^{1/\overline{z}} ,
\label{GL.7}
\ee
with a disorder-dependent exponent $\overline{z}$. This growth law applies across 
the accessible time region, and there is no hint of a crossover. 

A possible explanation of the numerical findings has been put forward by Paul et al.~\cite{ppr04} 
on the basis of a logarithmic 
dependence of $E_B$ on $L$, in place of the power law in Eq.~(\ref{GL.2hh}). However, 
Cugliandolo et al.~\cite{Cuglia,Iguain} have argued that the observation of a behavior 
consistent with Eq.~(\ref{GL.7}) does not necessarily rule out a growth law 
obeying Eqs.~(\ref{GL.4}) and~(\ref{GL.3}), since a disorder-dependent algebraic growth is 
compatible with Eq.~(\ref{GL.3}) as an intermediate regime, as it had been previously remarked
by Bouchaud et al.~\cite{Bouchaud} in the context of spin glasses.
Thus, if we rewrite Eq.~(\ref{GL.3}) as
\be
L^{\overline{z}} \simeq L^z e^{(L/\lambda)^{\psi}} ,
\label{GL.71}
\ee
we obtain 
\be
\overline{z} \simeq z + {1 \over \lambda^{\psi}}{L^{\psi} \over \ln L} 
\label{GL.72}
\ee
and, if $L^{\psi} /\ln L$ can be treated as a constant $c$ over the time interval of interest, 
the effective exponent is given by
\be
\overline{z} \simeq z + {c \over \lambda^{\psi}} = z + \frac{c\kappa}{T}. 
\label{GL.74}
\ee
The present understanding of the observed algebraic growth~(\ref{GL.7})
is to a sort of standstill between these two mutually excluding explanations.

In order to settle the issue, we have performed the most comprehensive simulations, to date, 
of coarsening in the $d=2$ Glauber-RBIM. Our simulations have been done on $N^2$ square lattices up to 
maximum time $t_m$ Monte Carlo steps (MCS). Taking a random configuration of up and down spins,
which mimics the disordered $T=\infty$ state before the quench, as initial condition for a run,
we have let the system evolve with
the updating rule which aligns spins with probability $1$ with the
Weiss field, if there is a majority of nearest neighbours pointing in the same direction,
while the usual Glauber rule is used when the Weiss field is zero. This is equivalent to take
couplings of the form $J_{ij} = J + \delta_{ij}$, with $\delta_{ij}$ uniformly distributed 
in $(-\epsilon, \epsilon)$ and then letting $T \rightarrow 0$ and $\epsilon \rightarrow 0$,
while keeping $\epsilon / T$ finite. Specifically, we have studied the coarsening 
dynamics with values  $\epsilon/T=0,0.5,1,1.5,2,2.5$.
Kinetics in this form is appropriate for deep quenches, where it provides
a considerable gain in the speed of computation, since updating trials are restricted to spins
not aligned with the Weiss field, whose number 
roughly scales like $1/L(t)$. In the above limit disorder and temperature enter the transition
rates in the form $\epsilon/T$ and, therefore, also results depend on disorder and temperature only through
the ratio $\epsilon/T$. Finally, we have checked that there is no difference
with the growth law and the aging properties of two-time quantities computed with
the standard Monte Carlo algorithm.

All statistical quantities have been obtained as an average over $N_{\rm run}$ 
independent runs. For each run, the system has a different initial condition and disorder configuration.
The values of $N, t_m$ and $N_{\rm run}$ for different values of $\epsilon/T$ are listed in Table~\ref{stat}.
\begin{table}
\begin{center}
  \begin{tabular}{|c|c|c|c|}
    \hline
Disorder values & System size & Maximum time & Number of runs \\ 
$\epsilon/T$     & $N^2$ & $t_m$ (MCS) & $N_{\rm run}$ \\ \hline 
$0.0$ & $12000^2$ & $1.01027 \times 10^5$ & $20$  \\ \hline
$0.5$ & $8192^2 $ & $3.12973 \times 10^5$ & $20$  \\ \hline
$1.0$ & $4096^2$ & $1.006735 \times 10^6$ & $75$  \\ \hline
$1.5$ & $4096^2$ & $1.006735 \times 10^6$ & $75$  \\ \hline
$2.0$ & $4096^2$ & $1.006735 \times 10^6$ & $75$  \\ \hline
$2.5$ & $4096^2$ & $1.006735 \times 10^6$ & $75$  \\ \hline
\end{tabular}
\end{center}
\caption{Numerical parameters for $d=2$ RBIM simulations for various disorder strengths.}
\label{stat}
\end{table}
Notice that we have taken huge system sizes for $\epsilon/T=0$ and $\epsilon/T=0.5$. This is necessary to obtain 
a reasonable time-window of domain growth before we encounter finite-size effects, as coarsening is more 
rapid for pure and near-pure systems with nonconserved dynamics. We have performed several checks to ensure that 
the data presented here are free of finite-size effects. This is crucial as the crossover phenomena we are 
investigating are subtle and are expected to occur at very late times. We also stress that the $t_m$-values 
for the present data sets are an order of magnitude (or more) larger than those of Paul et al.~\cite{ppr04}.

Our findings for the growth law, the autocorrelation function and the ZFCS are presented in the following subsections.

\subsection{Growth Law}

\begin{figure}
   \centering
  \rotatebox{0}{\resizebox{.85\textwidth}{!}{\includegraphics{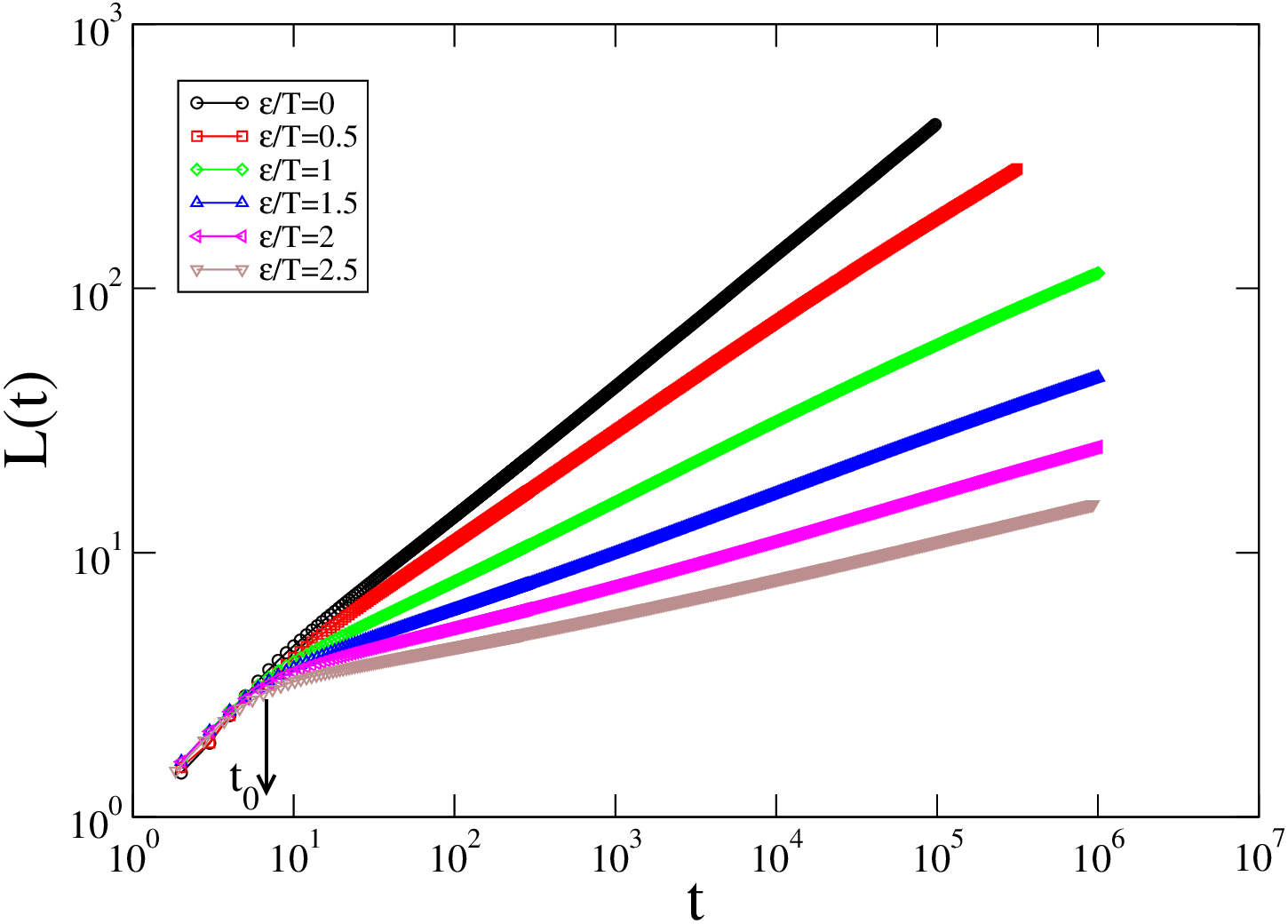}}}
  \vspace{0.5cm}
   \caption{Growth law in the pure case (top curve) and for various disorder strengths.}
\vspace{1cm}
\label{fig1}
\end{figure}

\begin{figure}
\centering
  \rotatebox{0}{\resizebox{.85\textwidth}{!}{\includegraphics{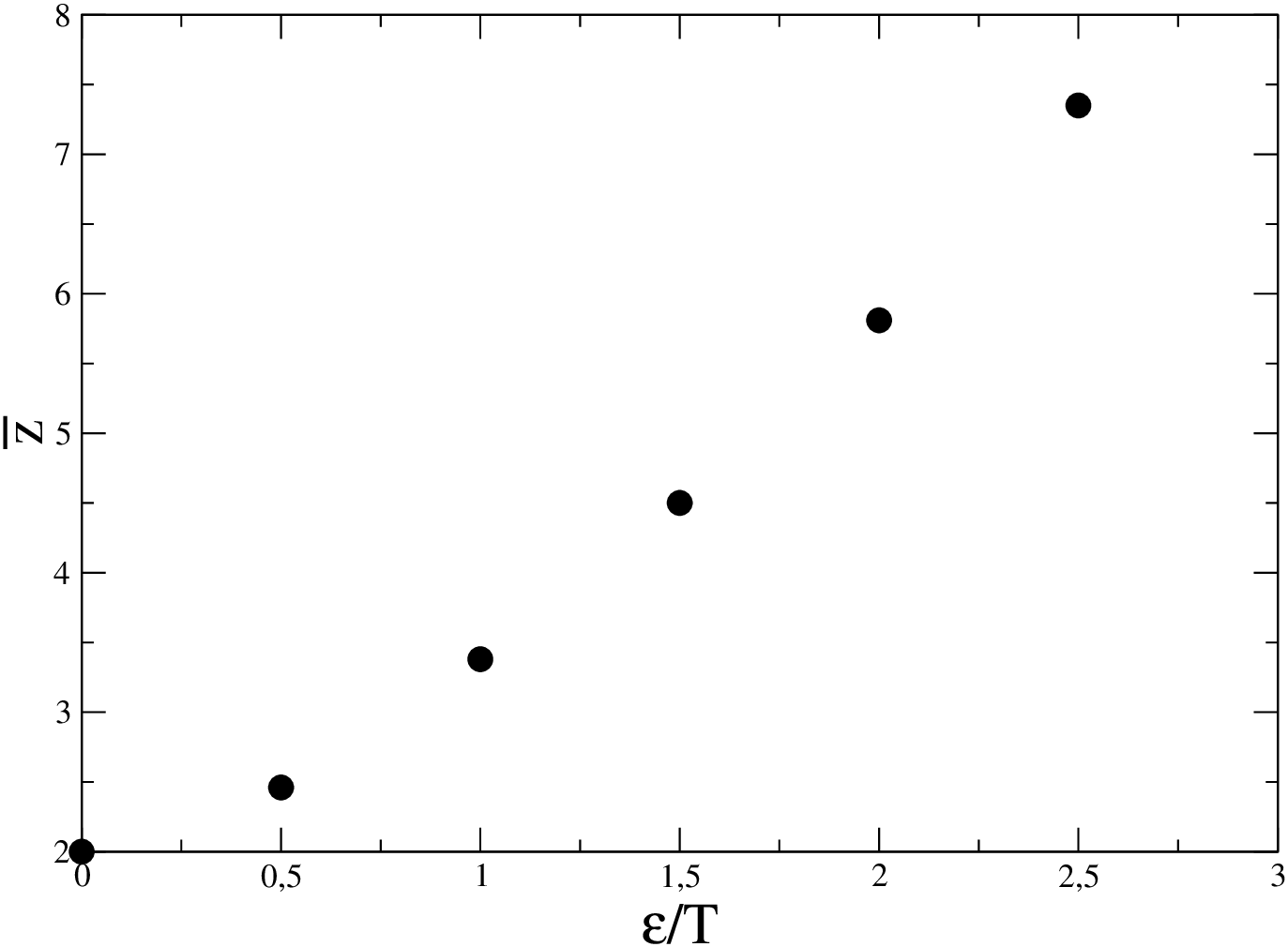}}}
  \vspace{0.5cm}
\caption{Disorder-dependent exponent $\overline{z}$ vs. $\epsilon/T$.}
\label{fig2}
\end{figure}

We have obtained $L(t)$ from the inverse density of defects, which is measured by dividing the number 
of sites with at least one oppositely-aligned neighbor by the total number of 
sites\footnote{We have checked that this definition is in good agreement with the usual 
method of measuring $L(t)$ by the half-peak-width of the equal-time structure factor.}. 
The plot of $L$ vs. $t$ in Fig.~\ref{fig1} shows the existence of two time regimes, separated by a 
microscopic time $t_0$ of order $1$. In the early time regime, for $t < t_0$, there is no dependence 
on disorder and growth is fast. The defects seeded by the initial condition execute rapid ballistic 
motion toward the nearby local minima. For $t > t_0$, the data sets in Fig.~\ref{fig1} appear to confirm 
previous observations \cite{ppr04,HP1,HP2} of a late-time regime with algebraic growth as in Eq.~(\ref{GL.7}). 
Measuring slopes, the values of the disorder-dependent exponent $\overline{z}$ are 
displayed in Table~\ref{expon} and Fig.~\ref{fig2}, with an $\epsilon/T$ dependence of
the type observed in Ref. \cite{HP2}.

\begin{table}
\begin{center}
\begin{tabular}{|c|c|}   \hline
$\epsilon/T$  & $\overline{z}$ \\  \hline
$0$ & $2$ \\         \hline
$0.5$ & $2.46$ \\    \hline
$1.0$ & $3.38$ \\    \hline
$1.5$ & $4.50$ \\    \hline
$2.0$ & $5.81$  \\    \hline
$2.5$ & $7.35$  \\    \hline
\end{tabular}
\end{center}
\caption{Exponent $\overline{z}$ for various disorder strengths.}
\label{expon}
\end{table}

\begin{figure}
   \centering
  \rotatebox{0}{\resizebox{.85\textwidth}{!}{\includegraphics{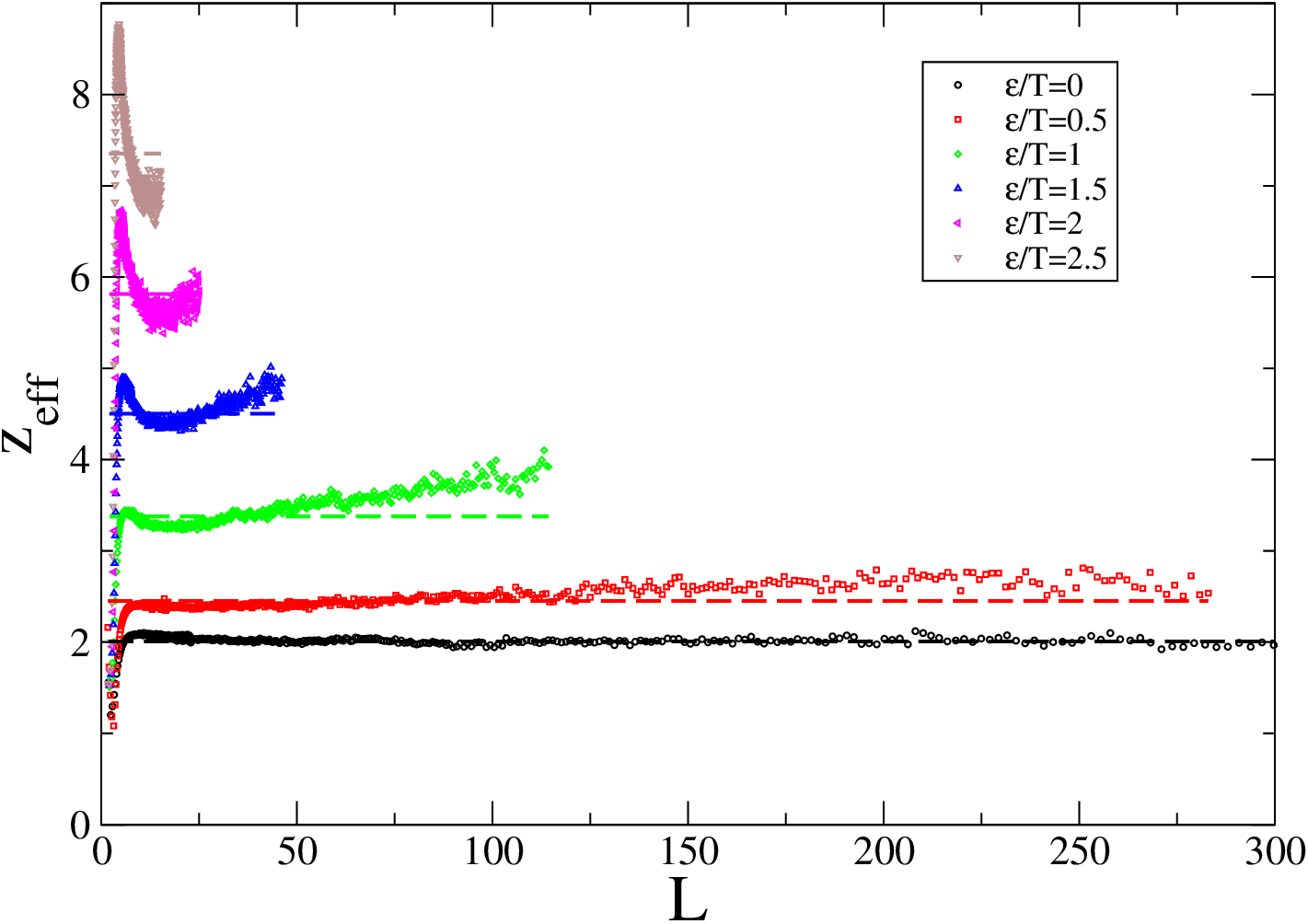}}}
  \vspace{0.5cm}
   \caption{Effective exponent $z_{\rm eff}$ vs. $L$. Straight dashed lines indicate $\overline{z}$.}
\vspace{1cm}
\label{fig4}
\end{figure}

The improvement with respect to previous work comes from the quality of our data, which allow 
for a more refined analysis. In addition to estimating $\overline{z}$ 
we have extracted the effective exponent $z_{\rm eff}(t)$ defined by
\be
{1 \over z_{\rm eff}(t)} = {d (\ln L) \over d (\ln t)} .
\label{GL.8}
\ee
In Fig.~\ref{fig4}, we have plotted $z_{\rm eff}(t)$ vs. $L(t)$ and here comes the main point in
the paper, since this
plot allows to discriminate between algebraic and logarithmic asymptotic growth
in favour of the latter choice. Indeed,
if growth was described 
correctly by Eq.~(\ref{GL.7}), then $z_{\rm eff}(L)$ ought to be independent of $L$, for $L \geq L_0=L(t_0)$,
with $z_{\rm eff}(L) \equiv \overline{z}$.
It is clear from Fig.~\ref{fig4} that this is the case only in the pure system ($\epsilon/T=0$). 
In all other cases, there is a slow time-increase which is not consistent with Eq.~(\ref{GL.7}). 
In order to make this feature clear, in Fig.~\ref{fig4} we have drawn the straight horizontal broken lines 
corresponding to the values of $\overline{z}$ from Table~\ref{expon}.
From the data follows i) that we can {\it exclude} the asymptotic validity of the algebraic growth law~(\ref{GL.7})
(and with it of the logarithmic dependence on $L$ of energy barriers) and ii) 
that the rather smooth and near-linear behavior of $z_{\rm eff}(L)$, for the 
smaller values of $\epsilon/T$ and $L \geq L_0$, is of the form
\be
z_{\rm eff}  = \zeta + b L^{\varphi} ,
\label{GL.9}
\ee
where $\zeta,b,\varphi$ are fit parameters.
Hence, inserting Eq.~(\ref{GL.9}) into Eq.~(\ref{GL.8}) 
and integrating over time from $t_0$ onward, we {\it derive} the growth law in the scaling form
\be
{t \over t_0} = {h(L/\lambda) \over h(L_0/\lambda)} ,
\label{12}
\ee
where
\be
h(x) = x^{\zeta} \exp(x^{\varphi}) 
\label{12bis}
\ee
and where we have set
\be
\lambda=(\varphi/b)^{1/\varphi}.
\label{14}
\ee 
We stress that in the above derivation there are no assumptions,
other than the form~(\ref{GL.9}) of $z_{\rm eff}$. Therefore, from the analysis of
the numerical data we predict the existence of a crossover from algebraic to logarithmic growth,
with the crossover length given by Eq.~(\ref{14}).

With the assumption $L_0 \ll \lambda$ (to be verified a posteriori), and 
using $h(x) \simeq x^{\zeta}$ for $x \ll 1$, Eq.~(\ref{12}) becomes
\be
(L/\lambda)^{\zeta}\,e^{(L/\lambda)^{\varphi}} = (L_0/\lambda)^{\zeta}\,t/t_0 
\label{13}
\ee
with the limiting behaviours
\be
L(t)  \simeq \left \{ \begin{array}{ll}
        D t^{1/\zeta}, \;\;$for$ \;\; L \ll \lambda , \\
        \lambda [\ln (t/t_0)]^{1/\varphi},  \;\; $for$ \;\; L \gg \lambda ,
        \end{array}
        \right .
        \label{15}
        \ee
where
\be
D = {L_0 \over t_0^{1/\zeta}}.
\label{16}
\ee
Comparing with Eqs.(\ref{GL.4}) and~(\ref{GL.3}) and identifying $\psi$ with $\varphi$,
we find that the observed growth is of the HH-type, except for the replacement of
the pure growth exponent $z$ with the disorder-dependent quantity $\zeta$
(which is distinct from $\overline{z}$). 
The top line of Eq.~(\ref{15}) accounts for the prior observations \cite{ppr04,HP1,HP2} of 
disorder-dependent power-law growth, without need of invoking the argument from Eq.~(\ref{GL.71}) up to
Eq.~(\ref{GL.74}). This is the pre-asymptotic regime followed 
by the asymptotically logarithmic one in the second line of Eq.~(\ref{15}). 
In addition, comparing Eq.~(\ref{GL.2bis}) with Eq.~(\ref{14}) we can make the 
identification
\be
\kappa = {Tb \over \varphi}.
\label{identif}
\ee
 
\begin{figure}
   \centering
  \rotatebox{0}{\resizebox{.85\textwidth}{!}{\includegraphics{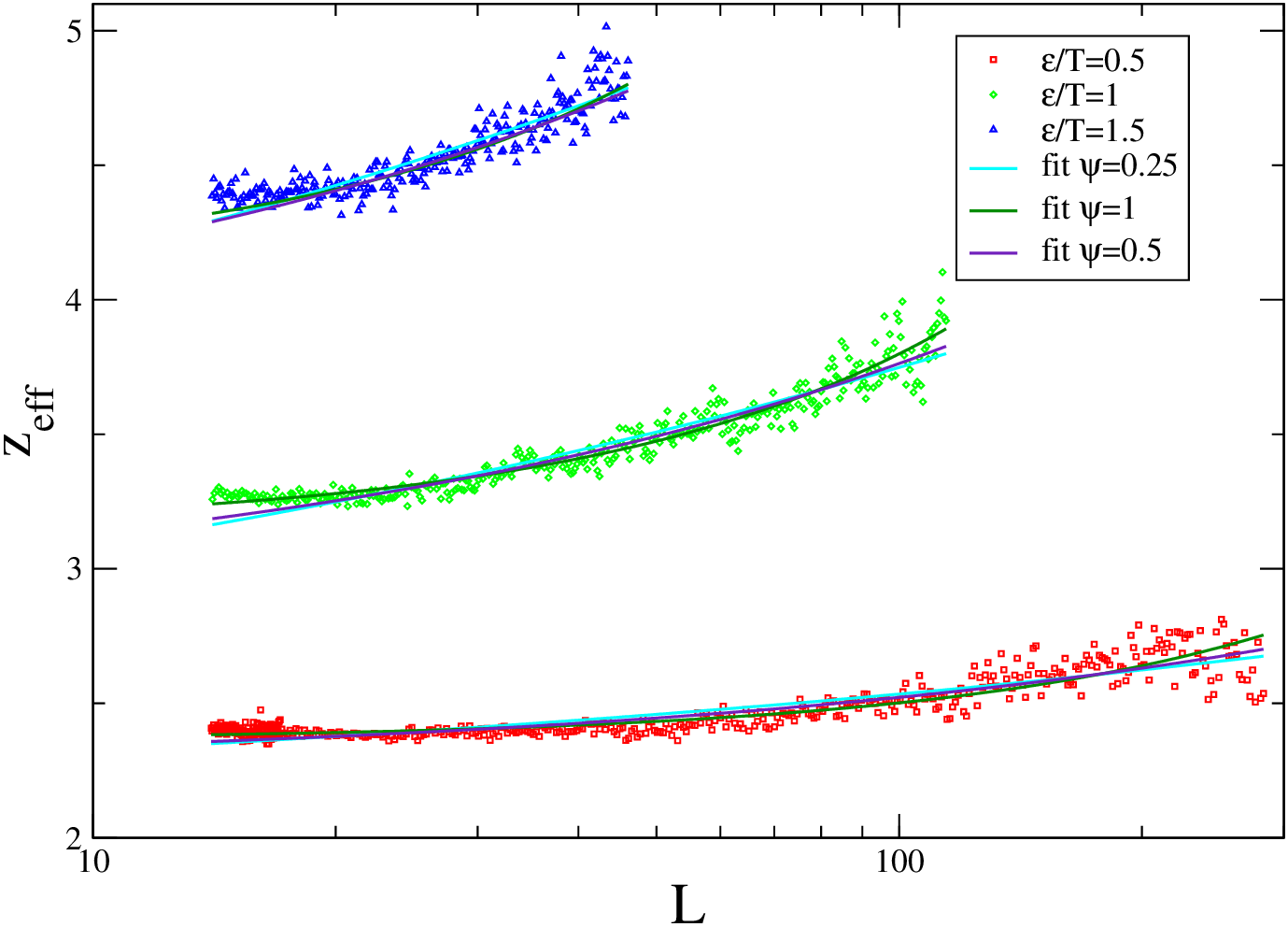}}}
  \vspace{0.5cm}
   \caption{Comparison of $z_{\rm eff}$ computed from Eq.~(\ref{GL.9}) with $\psi=1/4,1/2,1$
(continuous lines) and data from the simulations.}
\vspace{1cm}
\label{fig5}
\end{figure}

\begin{figure}
   \centering
  \rotatebox{0}{\resizebox{.85\textwidth}{!}{\includegraphics{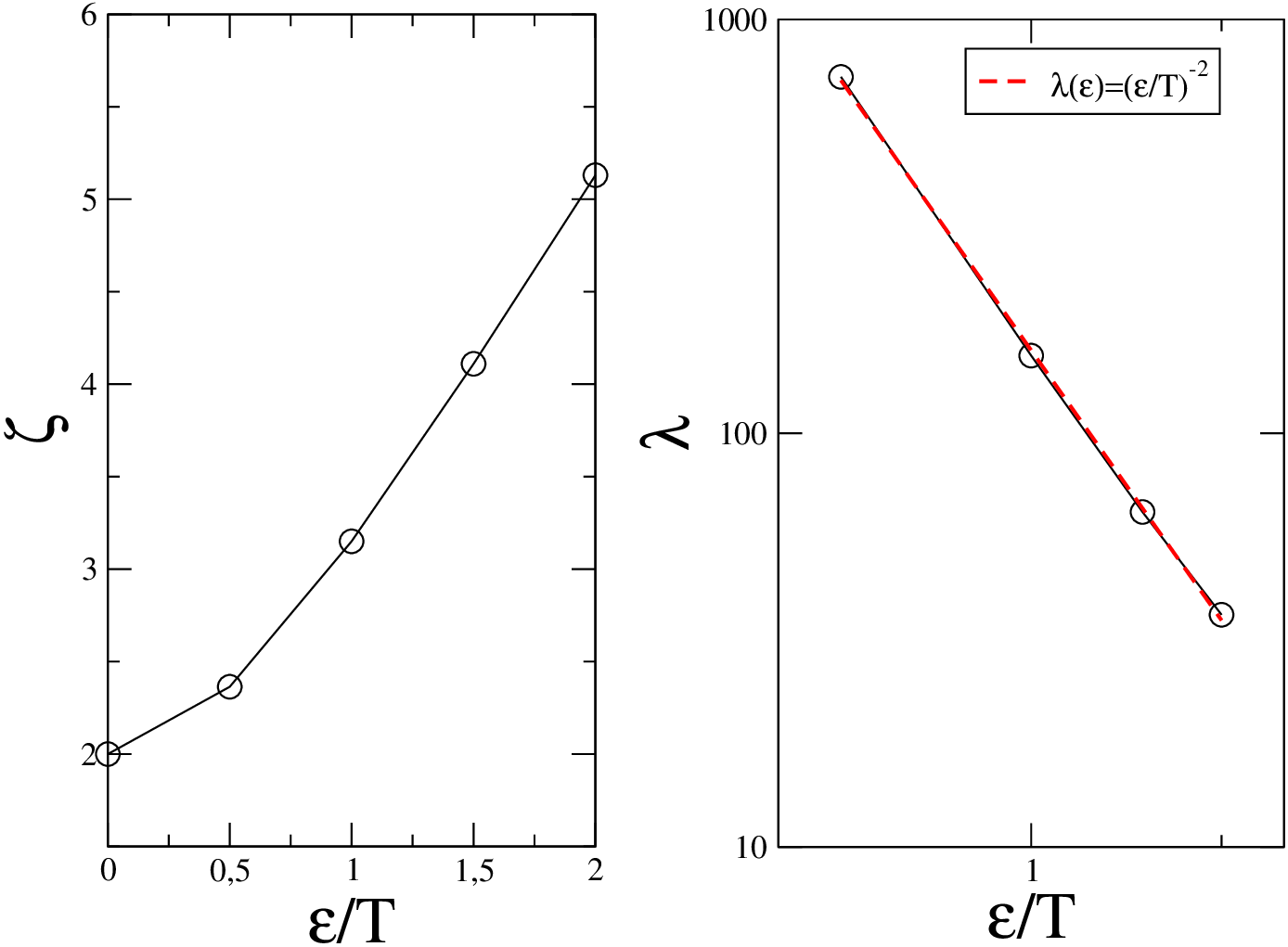}}}
  \vspace{0.5cm}
   \caption{Exponent $\zeta$ (left panel) and
$\lambda$ (right panel) vs. $\epsilon/T$ for $\psi=1$.}
\vspace{1cm}
\label{fig6}
\end{figure}

The first important conclusion, from the analysis of the numerical data presented above,
is the validation, at least at the qualitative level, of the HH scenario. 
Clearly, it would be important to check also on the quantitative HH prediction $\varphi=1/4$. 
However, it turns out to be difficult to obtain the numerical value of $\varphi$ from 
fitting the data sets in Fig.~\ref{fig4}. 
We have considered three trial values $\varphi =1/4, 1/2, 1$, and 
for each one of them we have fitted $\zeta,b$. In Fig.~\ref{fig5}, we have superposed these 
fits on the data sets for $z_{\rm eff}$ vs. $L$ for $\epsilon/T=0.5,1,1.5$. 
In the following we use $\varphi=1$, since this provides the best fit using the least-squares criterion. However,
differences are too small to commit to one particular value and we leave the issue open.
A precise determination of $\varphi$, as it is evident from Fig.~\ref{fig5}, requires to push
the accuracy of the numerical computation far beyond what we have done.
The corresponding values of $\zeta$ and $\lambda=1/b$ are listed in Table~\ref{abfit} and plotted in the
\begin{table}
\begin{center}
\begin{tabular}{|c|c|c|}   \hline
$\epsilon/T$  & $\lambda$ & $\zeta$ \\  \hline
$0.5$ & $762.2$ & $2.36$  \\    \hline
$1.0$ & $154.1$ & $3.15$ \\    \hline
$1.5$ & $64.5$ & $4.11$ \\    \hline
$2.0$ & $36.4$  & $5.13$  \\    \hline
\end{tabular}
\end{center}
\caption{Values of $\lambda$ and $\zeta$ for various disorder strengths.}
\label{abfit}
\end{table}
left and right panels of Fig.~\ref{fig6}. Notice that since $L_0 \sim 1$, as it is evident from Fig.~\ref{fig1},
the assumed inequality $L_0 \ll \lambda$ is satisfied. Furthermore,  $\lambda$ 
is found to decrease with increasing disorder as in Eq.~(\ref{EB.4bis}) with
\be
\lambda \sim (T/\epsilon)^2 .
\label{GL.10}
\ee
Hence, comparing with Eq.~(\ref{GL.2bis}), we obtain $\kappa \sim \epsilon^2/T$.

\begin{figure}
   \centering
  \rotatebox{0}{\resizebox{.85\textwidth}{!}{\includegraphics{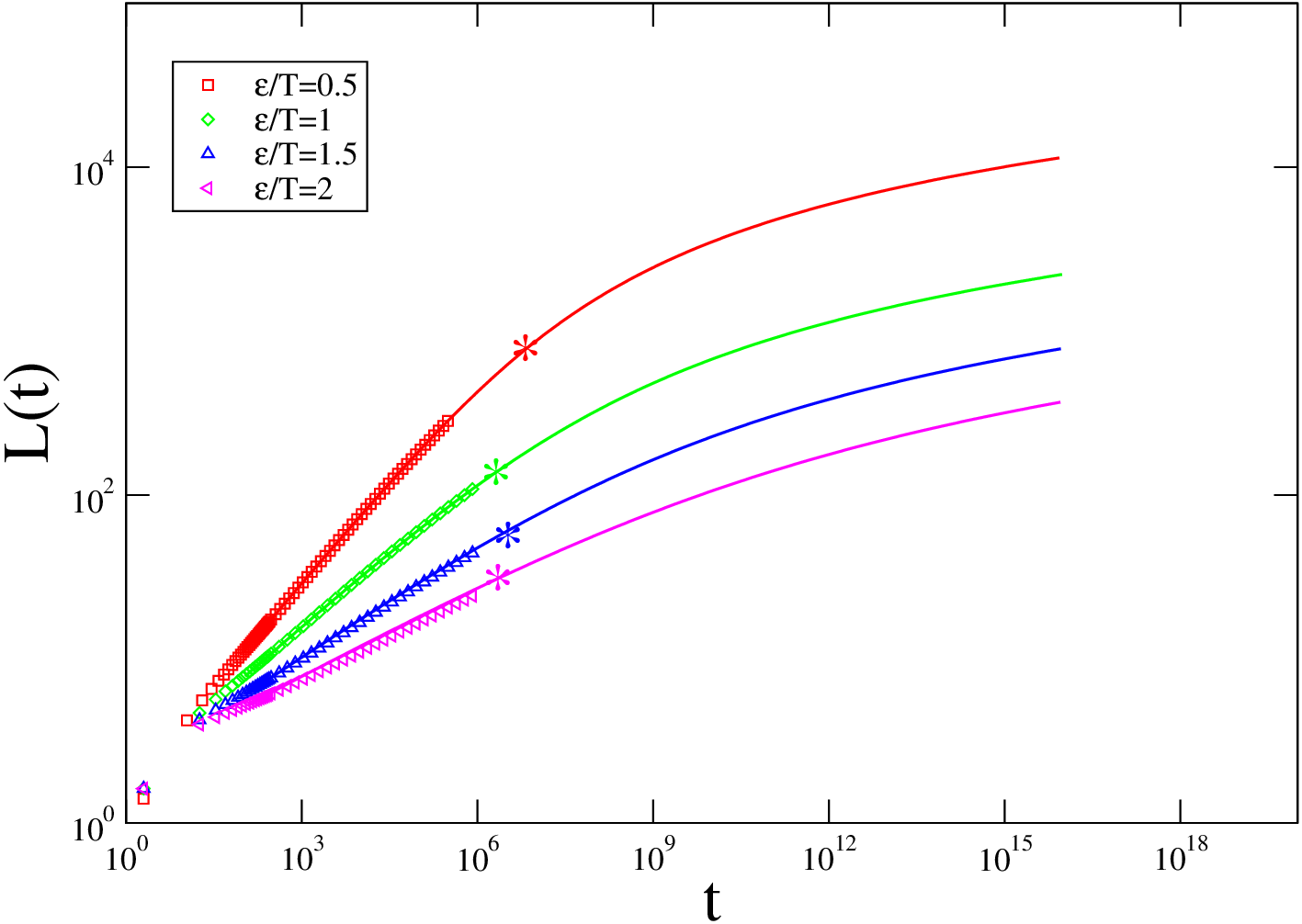}}}
  \vspace{0.5cm}
   \caption{Fully developed crossover obtained extrapolating the growth law from Eq.~(\ref{12}) (continuous lines) 
with parameters $\zeta$ and $\lambda$ from Table~\ref{abfit} and $\psi=1$. Symbols depict the numerical data
from the simulation. Asterisks mark the values of the crossover length $\lambda$.}
\vspace{1cm}
\label{fig7}
\end{figure}

Summarizing, our data for $z_{\rm eff}$ are evidence for the existence
of a crossover of which, however, we can access only the pre-asymptotic region,
since the values of the crossover length $\lambda$ are at the edge of the larger values of $L(t)$ reached in the
simulations for all disorder values (see Fig.~\ref{fig7}). 
Hence, numerically we can see only the onset
of the crossover. Nonetheless, a number of conclusions can be drawn.
As Eq.~(\ref{GL.10}) shows, disorder is relevant implying
the crossover to disorder-dominated behavior. 
The pre-asymptotic algebraic growth with a disorder-dependent 
exponent $\zeta$ signals that disorder, although being globally relevant, 
acts marginally in the neighborhood of the pure (unstable) fixed point.\footnote{The statements 
that the perturbation is marginal and that the pure fixed point is
unstable are compatible when nonlinear terms in the expansion around the fixed point 
are taken into account~\cite{wegner}.} As remarked previously, this happens also in the $d=1$ RBIM, where a 
disorder-dependent exponent arises in the pre-asymptotic regime, but the physical mechanism
producing these preasymptotic disorder-dependent exponents remains unclear. 
Finally, inserting the fitted values of the parameters into Eq.~(\ref{12}), the behavior
of the growth law can be analytically extrapolated for arbitrary large time. This has
been done in Fig.~\ref{fig7} where, in addition to finding excellent agreement in
the region where the simulation data are available, the overview of the predicted
crossover behavior is displayed.

\subsection{Autocorrelation Function}

\begin{figure}
   \centering
  \rotatebox{0}{\resizebox{.85\textwidth}{!}{\includegraphics{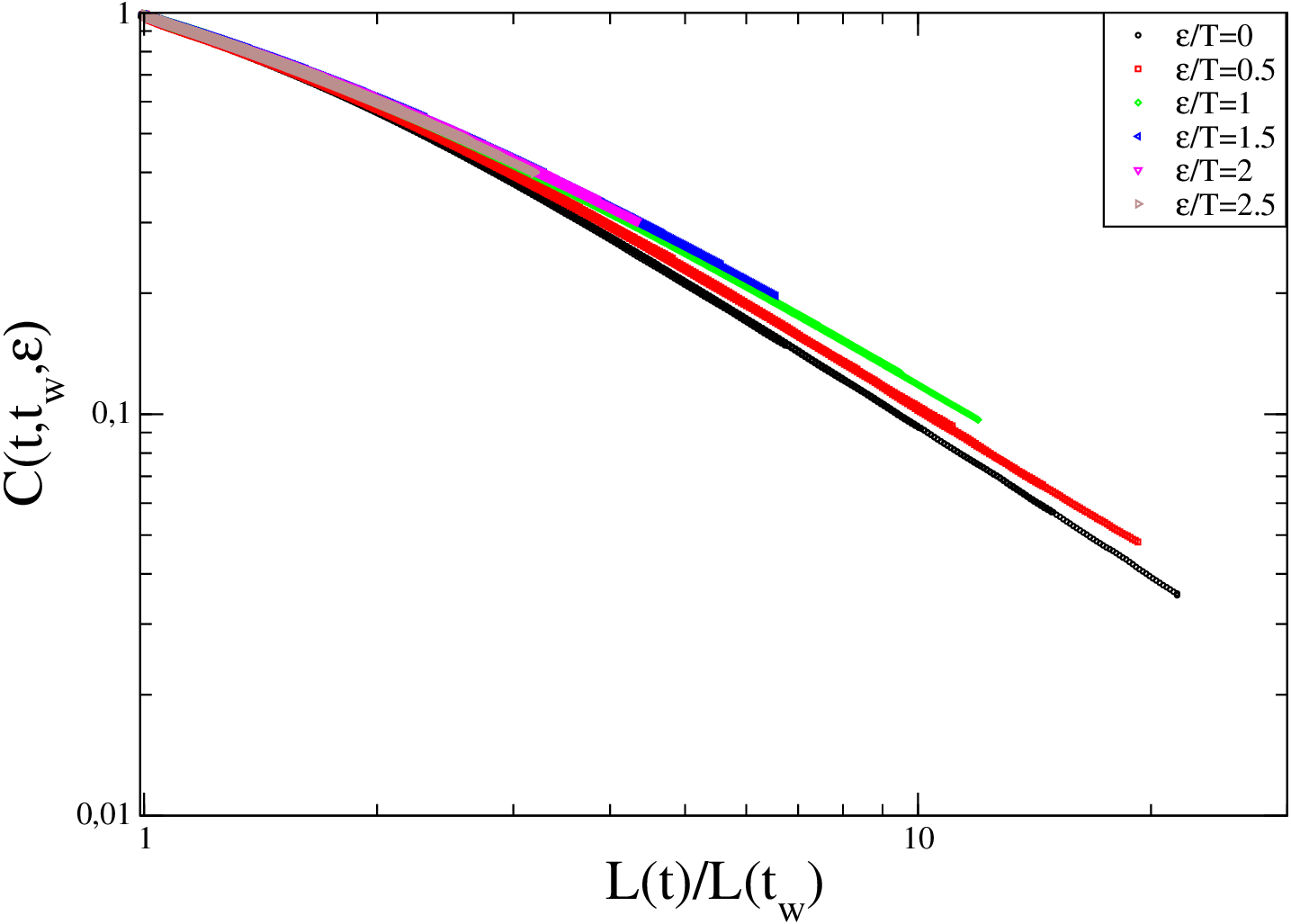}}}
  \vspace{0.5cm}
   \caption{Autocorrelation function for different disorder values. The 
$t_w$ values have been taken according to the rule
$t_w(k) = t_{\rm min} (t_m /2 t_{\rm min})^{(k-1)/10}$
where $t_{\rm min}=200$, $k$ is an integer running from $1$ to $11$ and $t_m$ is the maximum time
in Table~\ref{stat}. Accordingly, the waiting times are logarithmically spaced between $t_w=200$
and $t_w=t_m/2$.}
\vspace{1cm}
\label{fig8}
\end{figure}

Having established the existence of a crossover in the growth law, and that disorder is a relevant 
perturbation, we can anticipate that SU will not hold (see Sec.~\ref{GF}). If the autocorrelation 
function did obey SU as in Eq.~(\ref{IAF.1}), keeping in mind that $\alpha_C=0$ for pure and 
disordered systems, the plot of $C(t,t_w)$ against $L(t)/L(t_w)$ (for different values of $\epsilon/T$ and $t_w$) 
should collapse all the data sets on a single master curve. Figure~\ref{fig8} shows that this is
not the case. Instead, there is collapse only when $t_w$ is varied keeping $\epsilon/T$ fixed, but
the master curves of these {\it partial collapses}, corresponding to the different values of $\epsilon/T$, 
are distinct. 
What we have called {\it partial collapse} has also been reported 
in the context of the diluted Ising ferromagnet~\cite{Park,schehr}, without reaching, however, a definite
conclusion on the SU validity.

The behavior in Fig.~\ref{fig8} can be accounted for on the basis of the two-variable 
scaling $C(t,t_w,\epsilon) = \widetilde{F}_C(x,y)$ in Eq.~(\ref{IAF.4}) with $\alpha_C=0$. 
The pure case $\widetilde{F}_C(x,\infty)=F_C(x)$ corresponds to the lower black curve, 
where the collapse is very good since all the dependence on $t_w$ takes place through $x$. 
However, in the presence of disorder, $y$ is finite. The remaining curves in the plot are 
explained considering that with the $t_w$-values used in the simulations, 
the variation of $y$ is small while keeping $\epsilon/T$ fixed.
This produces the quasi-collapse 
on the coloured curves, each one corresponding to a distinct value of $\epsilon/T$. 
But the variation of $y$ becomes sizable when $\epsilon/T$ (and thereby $\lambda$) is changed, 
producing the separation of the curves with different colors. The trend in Fig.~\ref{fig8} 
shows that $\widetilde{F}_C(x,y)$ increases as $y$ decreases, for fixed $x$. Hence, the 
bundle of curves generated by letting $y$ vary is bounded below by the pure scaling function $F_C(x)$, 
and must be bounded above by the numerically inaccessible scaling function $\widehat{F}_C(x)$ of 
the disorder-dominated fixed point. Exactly the same behavior is observed in the autocorrelation 
function of the $d=1$ RFIM~\cite{decandia}.

Within the framework of our findings, we conclude that previously reported observations 
of SU in the autocorrelation function for the $d=3$ RFIM~\cite{rao,aron} and the $d=2$ RBIM~\cite{HP2} 
are due to time-ranges and disorder values deep inside the 
pre-asymptotic regime, where it is not possible to detect the onset of the crossover.

\subsection{Autoresponse Function}

The qualitative difference between the autoresponse and autocorrelation functions is that 
the exponent $\alpha_{\chi}$ in the scaling relation
\be
\chi(t,t_w,\epsilon) = L^{-\alpha_{\chi}}_w \widetilde{F}_{\chi} (x,y) ,
\label{AURF.1}
\ee
is not constrained to vanish, like $\alpha_{C}$, by geometrical requirements. 
Hence, in principle, the crossover could take the most general form of Eq.~(\ref{crossover}) 
with $\widehat{\alpha}_{\chi} \neq \alpha_{\chi}$.

Before analysing the data it is useful to summarize the structure of scaling in the pure case~\cite{CLZ}, where
Eq.~(\ref{AURF.1}) reads 
\be
\chi(t,t_w) = L^{-\alpha_{\chi}}_w F_{\chi} (x)
\label{AURF.1bis}
\ee
and for large $x$ the scaling function decays with the same exponent appearing in the prefactor
\be
F_{\chi} (x) \sim x^{-\alpha_{\chi}}.
\label{AURF.1tris}
\ee
As a matter of fact, the measurement of the decaying slope of $\chi(x,t_w)$ for large  $x$
is the fast numerical method to extract the scaling exponent $\alpha_{\chi}$.
The physical origin of this feature can be understood introducing the effective response associated with 
a single interface\footnote{For instance, a configuration with a single 
interface has been introduced in Eq.~(\ref{1d.1}).}, which is defined by~\cite{single}
\be
\chi(t,t_w) = \rho(t)\chi_{\rm eff}(t,t_w) 
\label{sing}
\ee
where $\rho(t)=L^{-1}(t)$ is the interface density. Then, from the scaling form in Eq.~(\ref{AURF.1bis}) follows
\be
\chi_{\rm eff}(t,t_w) \sim L_w^{(1-\alpha_{\chi})} xF_{\chi} (x) .
\label{sing.1}
\ee
On the other hand, if there is only one interface in the system the response is not expected to age,
yielding a time-translation invariant $\chi_{\rm eff}(t-t_w)$. Therefore, regarding the left hand side
as dependent on time trough $L(t-t_w)$ and considering the large $x$ regime where it is
possible to replace $L(t-t_w)$ by $L(t)$, the dependence on $L_w$ must drop out of the right hand side
implying $xF_{\chi} (x) \sim x^{(1-\alpha_{\chi})}$. This, in turn, implies the validity of Eq.~(\ref{AURF.1tris})
and
\be
\chi_{\rm eff}(t,t_w) \sim L^{1-\alpha_{\chi}}.
\label{sing.2}
\ee

\begin{figure}
   \centering
  \rotatebox{0}{\resizebox{.85\textwidth}{!}{\includegraphics{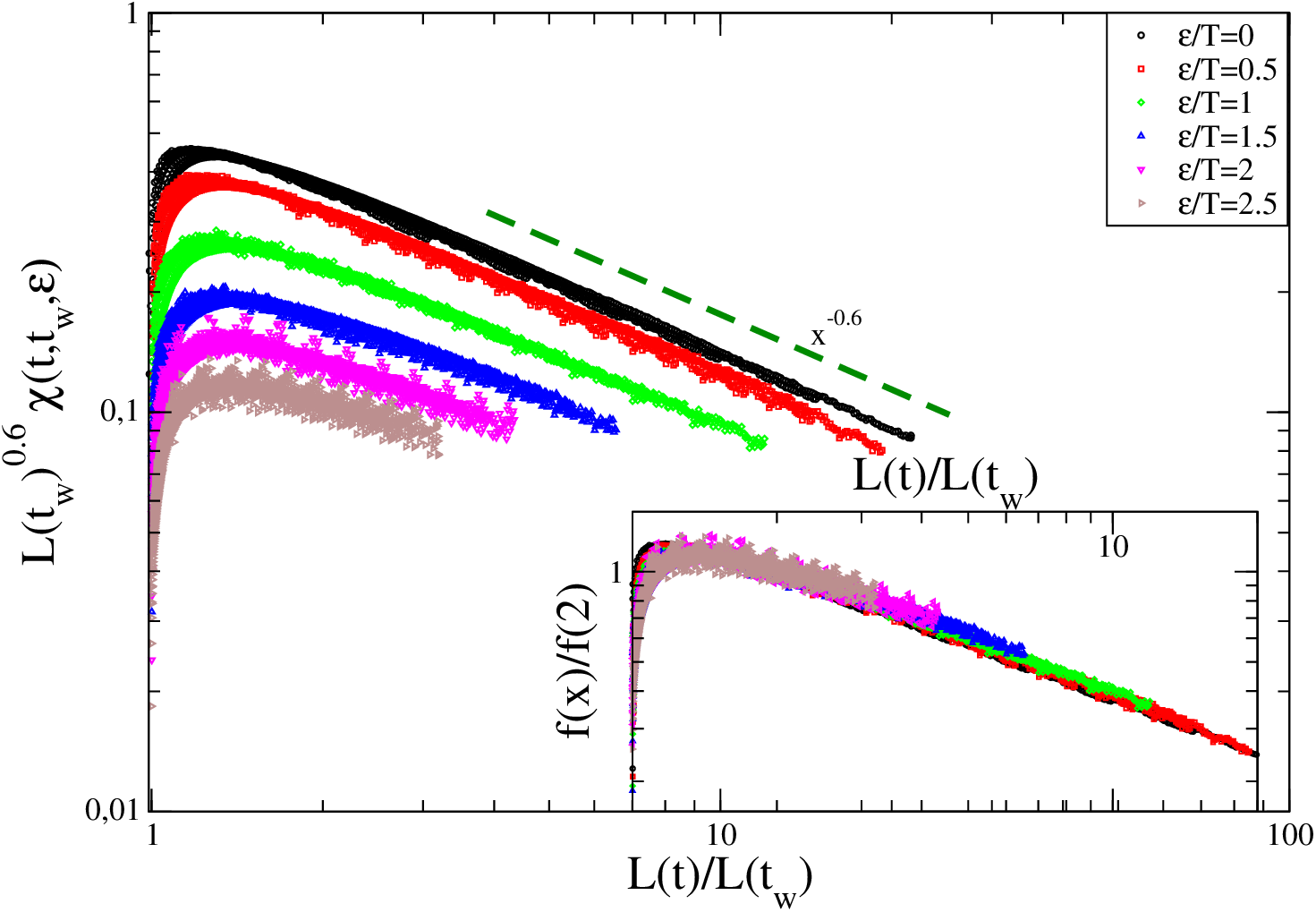}}}
  \vspace{0.5cm}
   \caption{Zero-field-cooled susceptibility for different disorder values and $t_w$ values
chosen according to the rule in the caption of Fig.~\ref{fig8}.}
\vspace{1cm}
\label{fig9}
\end{figure}

Let us now go to the data in presence of disorder. We have computed $\chi(t,t_w,\epsilon)$ with the algorithm
derived in Ref.~\cite{zerofield}, which does not require the
switching on of the external perturbation, thus improving the precision and the efficiency
of the numerical computation. The use of algorithms of this type is indispensable for the study
of linear response when the system is quenched to $T=0$. 
The quantity 
$L^{\alpha_{\chi}}_w \chi(t,t_w,\epsilon)$ has been plotted in Fig.~\ref{fig9} against 
$L/L_w$ for different values of $\epsilon/T$ and $t_w$, 
using for all disorder values $\alpha_{\chi}=0.6$, which is the value producing the best data collapse in the 
pure case (top black curves). The structure of the plot follows the same pattern as in Fig.~\ref{fig8}, 
with good partial collapse when $\epsilon/T$ is kept fixed and distinct master curves when $\epsilon/T$ is varied,
for the same reasons explained in the discussion of the autocorrelation function.
The collapse of
the curves with the pure exponent $\alpha_{\chi}$ shows clearly that
the numerical data belong to the preasymptotic region controlled by the pure fixed point up
to the onset of the crossover, since the $y$ dependence of the scaling function
$\widetilde{F}(x,y)$ is manifest in Fig.~\ref{fig9}.

Furthermore, since the different curves in the plot appear to be obtained one from the other
by a vertical shift,
in this region of parameters
the dependence on $x$ and $y$ can be taken to be of the multiplicative type
\be
\widetilde{F}(x,y) = f(x)g(y)
\label{AURF.6}
\ee
with $f(x)$ decaying algebraically for large $x$, exactly as in Eq.~(\ref{AURF.1tris}) for the
pure case. In order to highlight this feature, in the inset of Fig.~\ref{fig9} we have plotted
$\widetilde{F}(x,y) / \widetilde{F}(x=2,y)$ obtaining the superposition of the different
curves, that is the elimination of the $y$-dependence, which is a nice check on the validity of the
assumption made in Eq.~(\ref{AURF.6}). Therefore, the curve in the inset is the normalized plot of $f(x)$, whose 
dacay in agreement with Eq.~(\ref{AURF.1tris}) is a confirmation i) that scaling is controlled by the pure fixed point
and ii) of the validity of the single interface structure of the response.

\begin{figure}
   \centering
  \rotatebox{0}{\resizebox{.85\textwidth}{!}{\includegraphics{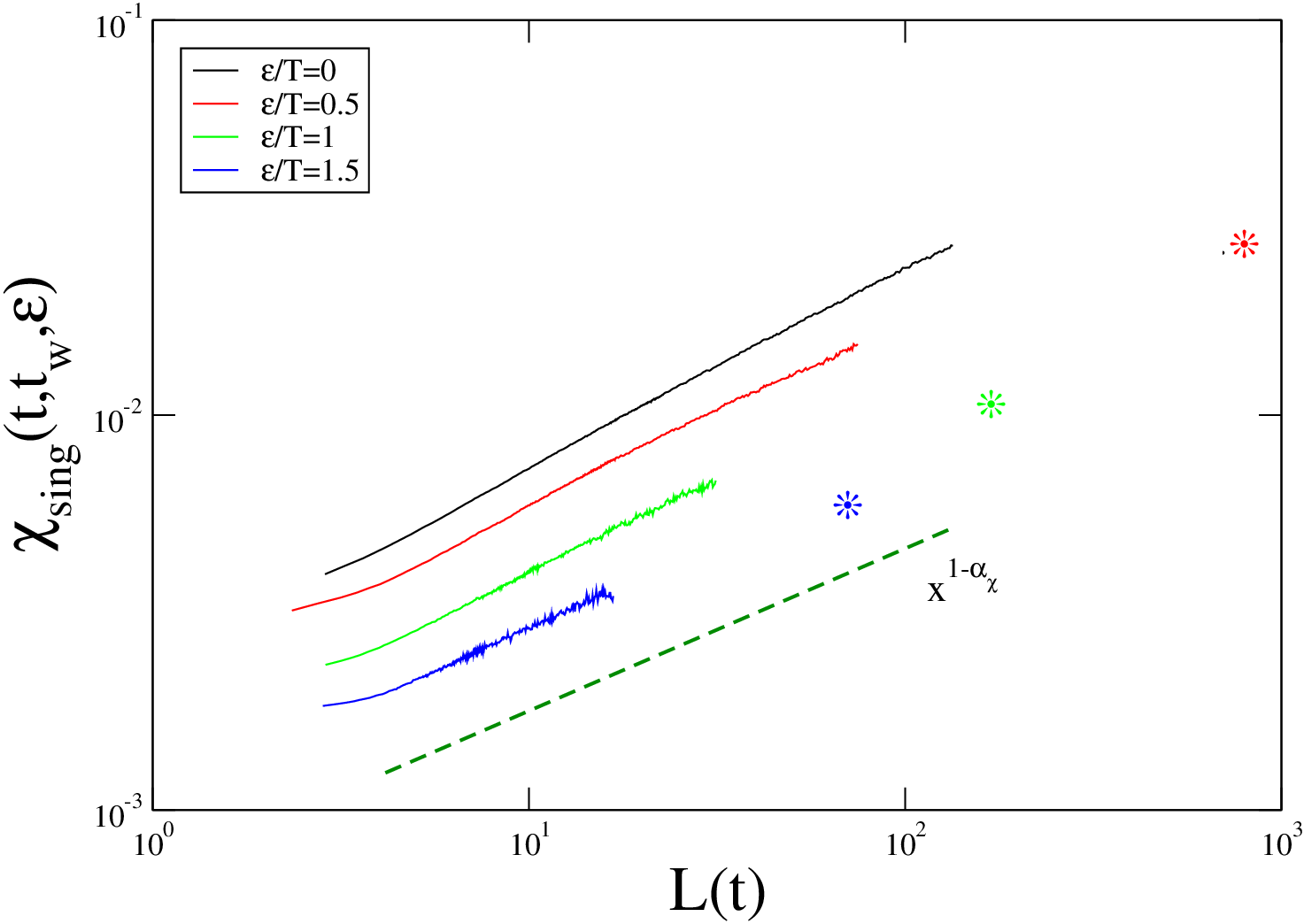}}}
  \vspace{0.5cm}
   \caption{Zero-field-cooled susceptibility for a single interface vs. L(t), taken from the data of
Fig.~\ref{fig1}, for different disorder values. Asterisks mark the values of the crossover length $\lambda$.}
\vspace{1cm}
\label{fig10}
\end{figure}

As a further check of this latter point,
we have measured the ZFCS $\chi_{\rm single}(t,t_w,\epsilon)$ after preparing an initial state 
with a single interface\footnote{The initial single interface is prepared along a diagonal
in the lattice, in such a way that spins on the interface experience zero Weiss field.
If the interface was prepared along the horizontal or vertical axis, the Weiss field
would not be zero and no flip could take place with the accelerated algorithm presented
at the beginning of this section.}
and for different disorder values. 
The plot in Fig.~\ref{fig10}, which also includes the pure case, does indeed show that $\chi_{\rm single}(t,t_w,\epsilon)$ 
grows with a power law in agreement with Eq.~(\ref{sing.2}). The exponent does not seem to
depend on $\epsilon/T$ and it is consistent with the expected value $1-\alpha_{\chi} \simeq 0.4$. Conversely,
in front of the power there is a prefactor $A$ dependent on   
$\epsilon/T$ (see Table~\ref{II}), which enters into the factor $g(y)$, in Eq.~(\ref{AURF.6}), 
and is responsible for the separation of the master curves in Fig.~\ref{fig9}.

\begin{table}
\begin{center}
\begin{tabular}{|c|c|}   \hline
$\epsilon/T$  & $A$ \\  \hline
$0.0$ & $0.00235$   \\    \hline
$0.5$ & $0.00184$   \\    \hline
$1.0$ & $0.00125$  \\    \hline
$1.5$ & $0.00093$ \\    \hline
\end{tabular}
\end{center}
\caption{Amplitude $A$ of $\chi_{\rm single}(t,t_w)$ extracted from the data of Fig.~\ref{fig10}
for different disorder values.}
\label{II}
\end{table}

In summary, since our data do not allow to access the full crossover region, we cannot make
statements on the interesting question whether the exponent 
$\widehat{\alpha}_{\chi}$ will be different or not from $\alpha_{\chi}$ when the disorder-dominated
asymptotics are reached.
However, precisely because what we observe is the pre-asymptotic regime controlled by
the pure fixed point, this result is important in itself, as explained in detail in Ref.~\cite{EPL},
since it brings new and independent evidence on the value of $\alpha_{\chi}$ in the pure system.
In fact, together with the $d=1$ result in Eq.~(\ref{alpha}), this result substantiates the conjecture 
put forward~\cite{single} for the pure system that $\alpha_{\chi}$ depends on dimensionality according to
\be
\alpha_{\chi} = (d-1)/2 ,
\label{AURF.2}
\ee
when $d \leq 3$. The numerical finding $\alpha_{\chi} \simeq 0.6$ does not exactly coincide with 
the value $\alpha_{\chi} = 0.5$, predicted
by Eq.~(\ref{AURF.2}) for $d=2$. This means that either the phenomenological formula or 
the precision of the numerical computation still
need improvement. However, it seems clear that the alternative conjecture~\cite{Barrat,HPGL,HP0} predicting  
\be
\alpha_{\chi} = 1 ,
\label{AURF.3}
\ee
with no dependence on dimensionality~\cite{HPGL}, is ruled out by the data from the $d=1$ and $d=2$ RBIM. 
Preliminary results on this point have been presented in Ref.~\cite{EPL}.

\section{Summary and Discussion}
\label{conc}

Let us conclude this paper with a summary and discussion of the results presented here. 
We have reported results from the most comprehensive simulations to date of domain growth 
in the $d=2$ ferromagnetic RBIM with nonconserved Glauber kinetics. We undertook this 
computationally expensive study with a specific purpose in mind, viz., to obtain a clear 
understanding of possible crossovers in the domain growth law and other statistical properties 
of the evolution morphology. As regards the domain growth law, our results are incompatible with algebraic growth
scenario of Ref.~\cite{ppr04} and are compatible with 
a crossover from a pre-asymptotic regime having power-law growth with a disorder-dependent 
exponent to an asymptotic logarithmic regime [$L(t) \sim (\ln t)^{1/\varphi}$]. The pre-asymptotic regime is consistent with 
existing results for the RBIM~\cite{ppr04,HP1,HP2}. To the best of our knowledge,
the predictive power of our result for the
late-stage result is new and allow us to give numerical support to the HH scenatio. 
However, in spite of the huge numerical effort, we cannot conclusively 
determine the exponent $\varphi$ which characterizes the logarithmic regime. 
Clearly, more work is needed in this direction.

We use RG arguments to understand the crossover in the growth 
law in terms of the competition between a pure fixed point and a disordered fixed point. 
Within this framework the corresponding crossover in two-time quantities, like 
the autocorrelation function $C(t,t_w,\epsilon)$ and the autoresponse function $\chi (t,t_w,\epsilon)$,
is expected. 
The scaling function for $C(t,t_w,\epsilon)$ shows the onset of the crossover from a pure form to a 
disordered one. Even though the numerical data are limited to the pre-asymptotic region,
there is clear evidence that the SU hypothesis does 
not hold for two-time quantities. A similar behavior is observed in the data for $\chi (t,t_w,\epsilon)$,
where the crossover region is not entered deep enough for $\widetilde{F}_{\chi}(x,y)$ to develop
the $y$-singularity which eventually should change the scaling exponent 
from the pure value $\alpha_{\chi}$ to the disorder value $\widehat{\alpha}_{\chi}$. 
Nonetheless, the measurement of the pure exponent $\alpha_{\chi} \simeq 0.6$ is
quite an interesting result in itself, adding new evidence in support of the phenomenological
conjecture of Eq.~(\ref{AURF.2}).

At a more general level, the RG framework formulated in this paper, in the context of the
$d=2$ RBIM, proves to be comprehensive enough
to give also a natural explanation for the known results for the $d=1$ RFIM and $d=1$ RBIM.
As such, it is a good candidate for the understanding of coarsening in disordered systems,
whose predictive power could be tested at higher dimensionality or in other systems,
like the site diluted Ising model, where the crossover region perhaps is more easily accessible
than in the RBIM~\cite{Park}.

\vspace{5cm}

{\bf Acknowledgements -} FC, EL and MZ wish to thank Leticia Cugliandolo for very
useful conversations on the subject of this paper.
FC and MZ acknowledge financial support from PRIN 2007 {\it Statistical Physics of Strongly
Correlated Systems in Equilibrium and Out of Equilibrium: Exact Results and Field
Theory Methods} JHLPEZ .
MZ wishes to thank the Jawaharlal Nehru Institute of Advanced Study and the School of
Physical Sciences of the Jawaharlal Nehru University for hospitality and financial
support.

\end{document}